# Versatile Electronic Devices Based on $WSe_2/SnSe_2$ Vertical van der Waals Heterostructure

**Wei Li\*, Xiang Xiao, Huilong Xu**

Hisilicon Research Department, Huawei Technologies Co. Ltd., Shenzhen, China
liwei110@huawei.com

Abstract

Van der Waals heterostructures formed by stacking of various two-dimensional materials are promising in electronic applications. However, the performances of most reported electronic devices based on van der Waals heterostructures are far away from existing technologies'. Here we report high performance heterostructure devices based on vertically stacked tungsten diselenide and tin diselenide. Due to the unique band alignment and the atomic thickness of the material, both charge carrier transport and energy barrier can be effectively modulated by the applied electrical field. As a result, the heterostructure devices show superb characteristics, with a high current on/off ratio of $\sim 3 \times 10^8$, an average subthreshold slope of 126 mV/decade over five decades of current change due to band to band tunneling, an ultra-high rectification ratio of $\sim 3 \times 10^8$ and a current density of more than $10^4$ A/cm$^2$. Further, a small signal half wave rectifier circuit based on majority carrier transport dominated diode is successfully demonstrated, showing great potential in future high speed electronic applications.

Two-dimensional (2D) materials are promising candidates for electronic and optoelectronic applications on Post Moore Era owing to their various useful properties.[1-6] Thanks to the weak van der Waals (vdW) interactions between individual layers, different 2D materials can be vertically stacked to form hetero systems without the restriction of lattice mismatch and the need of complex material growth procedures required in conventional heterostructure fabrications.[7-9] The atomic level flatness and the dangling bond free surface of 2D materials facilitate the atomically sharp interface and minimal interface charge traps of the stacked vdW heterostructures (vdWHs), which are important for realizing high performance heterostructure devices.[8, 10-13] Furthermore, the big family of 2D materials offers a huge flexibility in engineering the proper energy band alignments and other electronic properties of the vdWHs by combining the desired 2D materials.[7-9, 14-17] Various vdWHs have been studied in the recent years, including tunneling field-effect transistors (TFETs),[15, 18-20] memory devices,[17, 21-23] vertical field-effect transistors (VFETs),[16, 24, 25] heterostructure diodes[14, 20, 26, 27] and optoelectronic devices.[28-31] However, among these results, most of the realized vdW heterostructure devices demonstrated a limited current density of less than $10^2$ A/cm$^2$ and rectification ratio (RR) of less than $10^7$ at room temperature (RT), which hinder the application of vdWH devices.[14, 16, 18, 20, 26, 27] In heterostructure devices, RR and on state current density mainly rely on the modulation of the Fermi level of the channel, the band offset between the materials and the serial resistance at the contacts. Tungsten diselenide (WSe$_2$) has an ambipolar behavior with a high carrier mobility, where the conductive characteristics can be effectively tuned by the applied electrical field.[4, 10, 32] Tin diselenide (SnSe$_2$) with a large electron affinity of ~ 5.1 eV and a moderate bandgap of ~1.0 eV is a degenerate n-doped semiconductor.[33,34] Multilayer-stacked WSe$_2$/SnSe$_2$ heterojunctions were predicted to be a near broken-gap band alignment.[33, 34] Thus, band to band tunneling (BTBT) can occur at a small electric field across the heterojunction, which is important for TFET and Zener diode applications.[34] Further, the large band offset between WSe$_2$ and SnSe$_2$ can facilitate the carrier transport in one direction while suppress the transport in the opposite direction with a large energy barrier at the interface. This is a prerequisite for a diode with a large rectification ratio.

In this article, we present high performance heterostructure devices based on vdW stacked WSe$_2$/SnSe$_2$ materials. Both static electrical doping and charge transfer doping techniques are introduced to modulate the electrical performances of the devices. Due to the near broken-gap band alignment of the WSe$_2$/SnSe$_2$ heterojunction and the bipolar conduction of WSe$_2$, the device can function as p-n junction forward diode, Zener diode, n-n junction backward diode, metal-oxide-semiconductor field-effect transistor (MOSFET) and TFET under certain bias conditions. As n-n backward diode, the device exhibits a high RR of 3 x $10^8$ and current density of ~ $10^4$ A/cm$^2$ at room temperature (RT), which are one order higher than previous reported semiconductor vdWHs.[16, 26] The vdWH device aslo shows a low average subthreshold slope (SS$_{ave}$) of ~ 126 mV/dec over 5 orders when acts as TFET and presents a high current on/off ratio of more than $10^8$ when functions as MOSFET. Furthermore, a small signal (With amplitude ~ 0.4 V) half wave rectifier circuit based on the backward diode is demonstrated for the first time in vdWHs.

**Device structure and basic characterizations**

A schematic of the cross section of the vdW stacked WSe$_2$/SnSe$_2$ heterostructure device is shown in Figure 1a. The WSe$_2$/SnSe$_2$ were mechanically transferred onto a heavily p-doped silicon substrate coated with atomic layer deposition (ALD) grown Al$_2$O$_3$. The detailed transfer process was described elsewhere.[35] Raman characterizations of WSe$_2$, SnSe$_2$ and their overlapped area (with SnSe2 as top layer) are shown in Figure 1b. The $E_{2g}$, $E_{2g}'$ and $A_{1g}$ modes of WSe$_2$ are positioned at ~ 138.4, 250.1 and 257.8 cm$^{-1}$, respectively.[36] For SnSe$_2$, two distinct peaks located at 112.4 and 187 cm$^{-1}$ are observed, which are regarded as $E_g$ and $A_{1g}$ modes, respectively.[37] The overlapped region inherits all the peaks from WSe$_2$ and SnSe$_2$ with a decreased intensity, which is due to the strong interlayer coupling between the two layers.[11] Shown in Figure 1c is the optical image of the fabricated devices. Multiple electrical contacts were fabricated on WSe$_2$ and SnSe$_2$. The thickness of the WSe$_2$ and SnSe$_2$ flakes are determined to be 7.2 nm and 39.1 nm by atomic force microscopy (AFM), respectively.

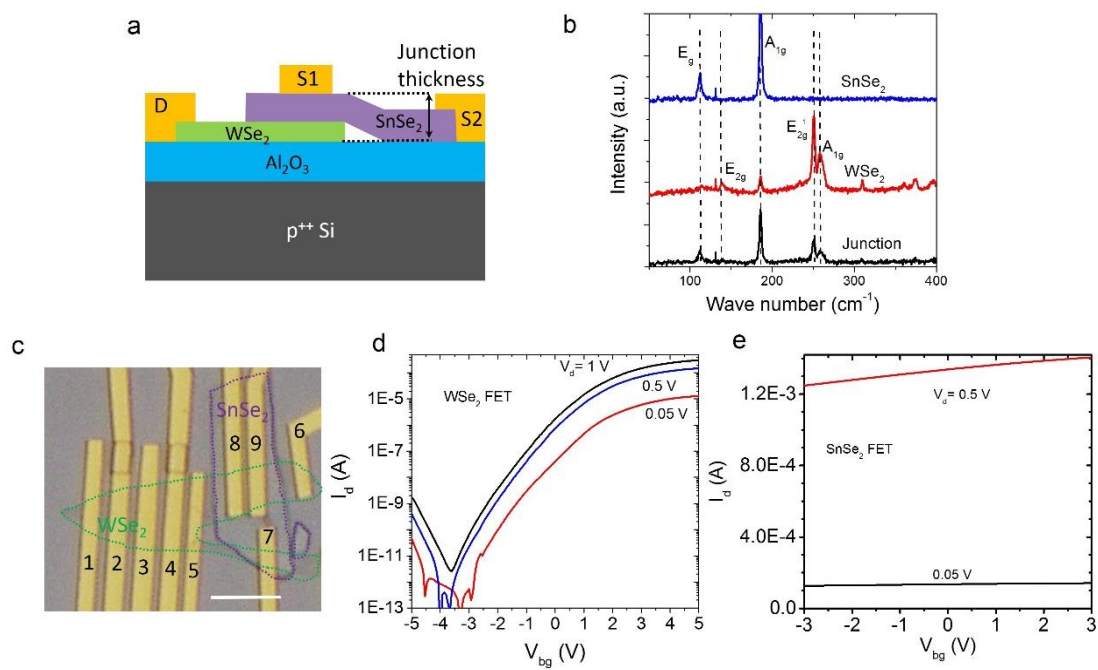

Figure 1. Device structure and characterizations. a, Schematic of cross section of the WSe$_2$/SnSe$_2$ heterostructure device. The substrate is ALD grown 10 nm thick Al$_2$O$_3$ on top of a heavily p doped silicon. The source (S1, S2) and drain (D) electrode are Pt/Au (8/60nm). b, Raman spectroscopy characterization of WSe$_2$, SnSe$_2$ and their overlapped area excited by 514 nm laser. c, Optical image of the devices. The edges of the WSe$_2$ and SnSe$_2$ flakes are marked by dotted lines with green and purple color, respectively. WSe$_2$ are contacted with electrode 1 to 6 and SnSe2 are contacted with electrode 7 to 9. Scale bar, 5 μm. d, Transfer characteristics of a WSe$_2$ FET (Electrodes 4 and 5). Channel length and width are 0.6 μm and 5.7 μm, respectively. e, Typical transfer characteristics of a SnSe$_2$ FET (Electrodes 8 and 9). Channel length and width are 0.6 μm and 10 μm, respectively.

The electrical characterizations of WSe$_2$ and SnSe$_2$ FETs are demonstrated in Figure 1d and e. The WSe$_2$ FET shows a n-type dominated ambipolar conduction behavior (Figure 1d). The extracted maximum field effect mobility of electrons are ~ 18 cm$^2$/Vs

(Supplementary Fig. 1). The transfer curves of WSe$_2$ FET shows a high current density around 50 μA/μm and a large on/off current ratio over $10^8$. These results are comparable with the best reported results.[4, 32] The increase of the off state leakage current and the decrease of the threshold voltage of the WSe$_2$ FET when drain voltage increases are due to the well-known gate induced drain leakage and drain induced barrier lowering effect, respectively.[38, 39] The SnSe$_2$ FET presents n-type behavior with a large drain current and weak gate modulation (Figure 1e), indicating SnSe$_2$ is a degenerate n-doped semiconductor.

**Heterostructure device transfer properties**

The typical transfer ($I_d$-$V_{bg}$) properties and band diagram descriptions of WSe$_2$/SnSe$_2$ heterostructure device are shown in Figure 2. To better understand the transport properties of the heterostructure, different bias conditions are explored. Here, we define the gate voltage where the drain current has a minimum value in each $I_d$-$V_{bg}$ curve as $V_{off}$. Shown in Figure 2a is the drain current as a function of the gate voltage under different drain biases. When biased at $V_d$ < 0 V, the device shows an current on/off ratio as high as 3 x $10^8$ under $V_d$= -2.5 V. Shown in Figure 2b is the current on/off ratio as a function of the drain bias. When the drain bias increases from – 3 V to 3 V, the on/off ratio decreases from ~$10^8$ to ~ $10^6$. This behavior is due to different carrier barriers at different bias conditions.

Shown in Figure 2c-f are the energy band diagrams of the heterostructure device under different gate and drain bias conditions. As mentioned before, WSe$_2$/SnSe$_2$ heterostructure has a near broken-gap band alignment.[33,34] When $V_{bg}$-$V_{off}$ < 0 V, the WSe$_2$ layer will be electrically p-type doped. Due to the electrical screening by WSe$_2$ and the degenerate n-type doping, the SnSe$_2$ layer remains n-type. Under positive drain bias, the holes from the drain side will be injected into WSe$_2$, while the electrons will be injected into SnSe$_2$ from the source side (Figure 2c). When the carriers reach the junction area, they may recombine (dashed arrow) or overcome (solid arrow) the junction barrier and then collected by the electrodes. Since the barrier height for the carriers at the interface are as high as around 1 eV, the current here might be dominated by recombination. The recombination dominated current are also observed in other vdWH systems.[20, 26] Under negative drain bias, the energy band of WSe$_2$ will bend upward. Carriers injected from the drain will be collected by the source through the band to band tunneling (BTBT) from the valence band of WSe$_2$ to the conduction band of SnSe$_2$ (Figure 2d). The current at positive drain bias is higher than the negative bias when negative gate voltages apllied, which is due to the larger |$V_{bg}$-$V_d$| for positive drain bias results more hole accumulation in the valence band of WSe$_2$, eventually leading a higher current.

When $V_{bg}$-$V_{off}$ > 0 V, the electrons will accumulate in WSe$_2$. Under $V_d$ > 0 V (Figure 2e), the majority carrier (electrons) from the source side will face a high and wide barrier at the junction interface, resulting in a low current. While $V_d$ < 0 V (Figure 2f), the electrons will tunnel through the thin Schottky barrier at the drain/WSe$_2$ interface and then meet a very small barrier at the interface, which results in an always higher current than the positive bias condition as shown in Figure 2a. For comparison, the current on/off ratio of the devices based on different vdWH systems are summarized in Figure 2g.[16, 24, 40] Due to the unique band alignment properties of WSe$_2$/SnSe$_2$ heterostructure, our results show a better on/off current ratio performance than previous reports.[16, 24, 40] The on/off ratio of the devices as a function of WSe$_2$ thickness are also shown in supplementary Figure 4.

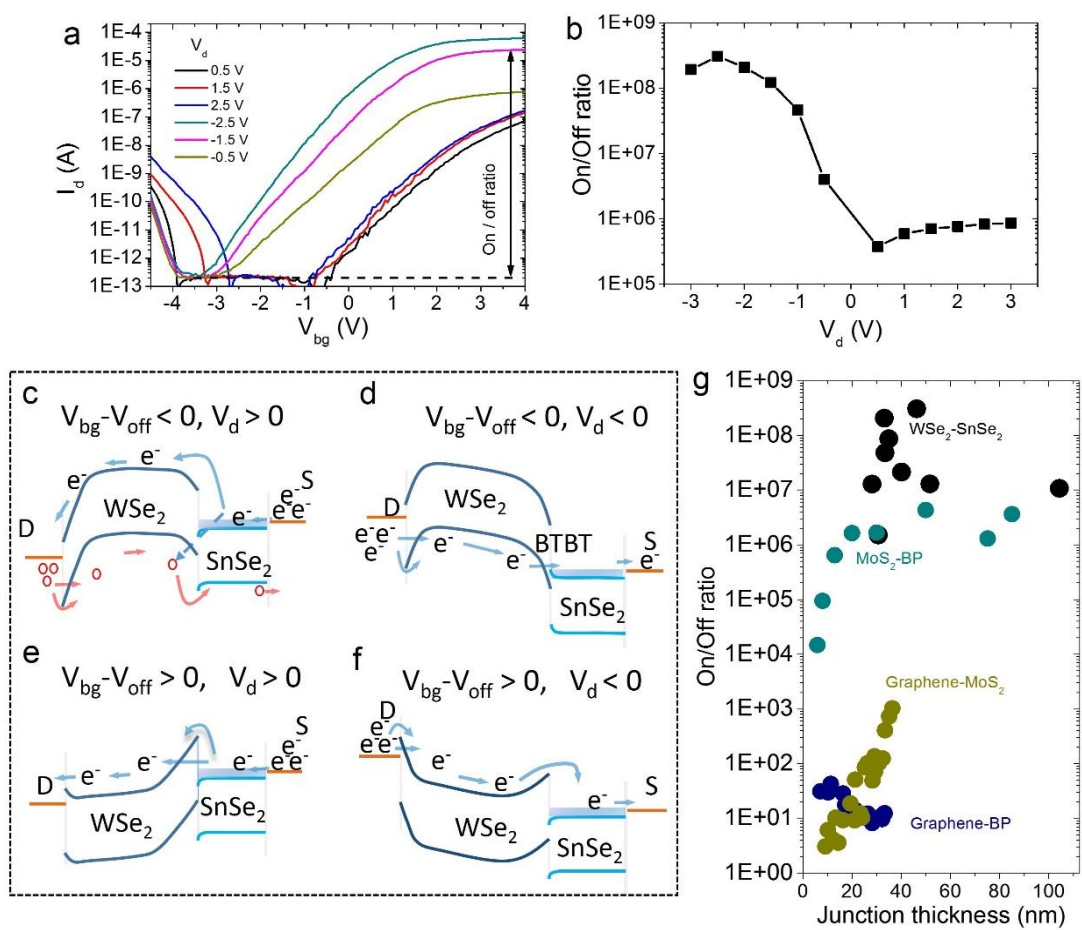

Figure 2. Transfer ($I_d$-$V_{bg}$) properties of the heterostructure device (Electrodes 6 and 9 in Figure 1c) and the related energy band diagrams. a, $I_d$-$V_{bg}$ transfer curves of WSe$_2$/SnSe$_2$ heterostructure device in semilogarithmic scale. b, Current on/off ratio at different drain voltages. c-f, Qualitative energy band diagrams of the heterostructure device at different bias conditions. g, Current on/off ratio comparison of heterostructure devices for different vdW heterostructure systems. The junction thickness definition is shown in Figure 1a. Reference data are from this work (WSe$_2$-SnSe$_2$), ref.[16] (MoS$_2$-BP), ref.[24] (Graphene-MoS$_2$) and ref.[40] (Graphene-BP), respectively.

**Heterostructure device output properties**

The output ($I_d$-$V_d$) properties of the heterostructure device and the rectifier circuit based on the device are illustrated in Figure 3. The semilogarithmic scale drain current as a function of drain voltage at different gate voltages is shown in Figure 3a. The band diagrams at different bias conditions can be similarly divided into four situations as described in previous section. When $V_{bg}$-$V_{off}$ < 0 V, at forward biases, the device becomes more conductive with the drain bias increases, since there are more accumulated hole carriers at the WSe$_2$ layer. While $V_{bg}$-$V_{off}$ > 0 V, WSe$_2$ is electrically n-type doped. At forward biases, the majority carriers (electrons) face a large barrier at the junction interface, which results in a small current. At reverse biases, the majority carriers will face a small barrier and results in a large current. Thus, the device resembles a nn backward diode behavior. When the $V_{bg}$ is shifting from -3 V to -1 V, the forward current is decreasing due to the decreasing of the hole concentration in the WSe$_2$ layer(Figure 3a solid red arrow). While the $V_{bg}$ is shifting from 1.5 V to 3 V, the energy band of WSe$_2$ will be severely bent downward, resulting a decreased barrier width at the junction interface. Thus, besides crossing the barrier by thermal injection, the electrons can also tunnel through the barrier. Therefore, an increased forward current is observed (Figure 3a dashed red arrow). The reverse current increases as $V_{bg}$ sweeping from -3 V to 3 V (Figure 3a solid black arrow), which is due to the decreases of the barrier width at the drain/WSe$_2$ interface (Figure 2f).

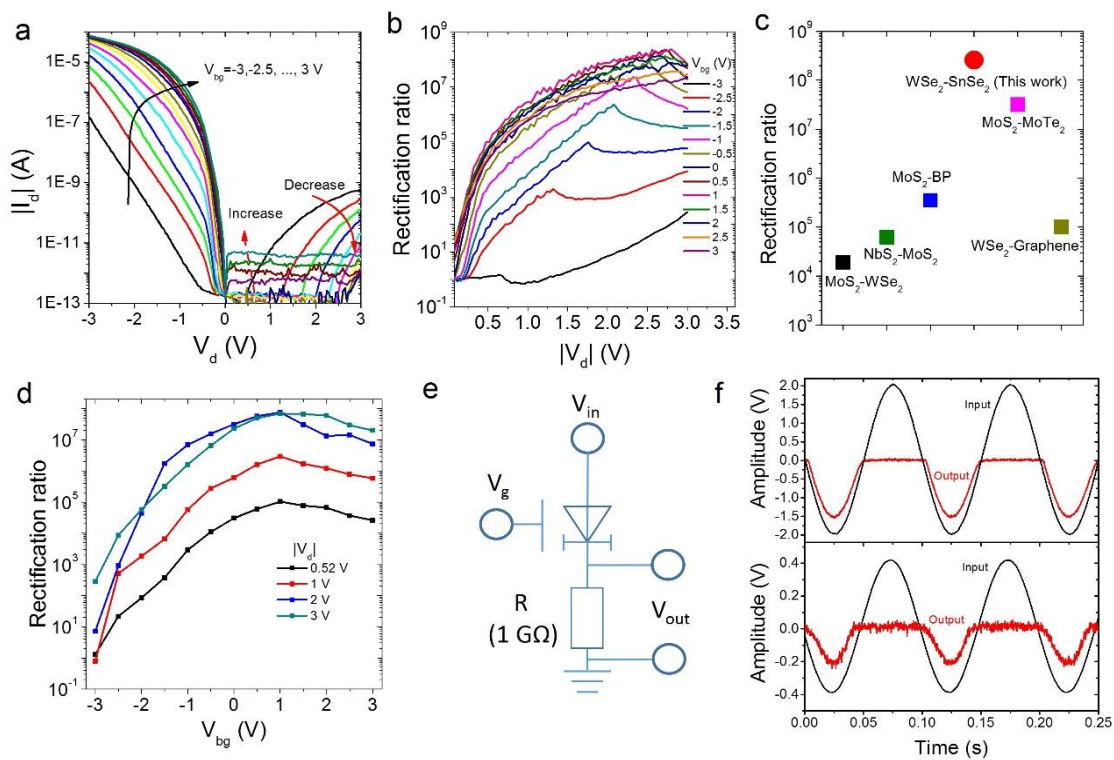

Figure 3. Output ($I_d$-$V_d$) properties of the WSe$_2$/SnSe$_2$ heterostructure device (Electrodes 6 and 9 in Figure 1c). a, $I_d$-$V_d$ curves of the device at different gate voltages from -3 V to 3 V at a step of 0.5 V in semilogarithmic scale. b, The rectification ratio (RR) of the diode as a function of drain bias at different gate voltages. RR is defined as |$I_{Reverse}$/$I_{Forward}$|, where $I_{Reverse}$ and $I_{Forward}$ are the reverse current at – $V_d$ and forward current at $V_d$, respectively. c, The maximum RR comparison of devices for different vdWH systems at RT. The reference data are from ref. [20] (MoS$_2$-WSe$_2$), ref. [27] (NbS$_2$-MoS$_2$), ref. [16] (MoS$_2$-BP), ref. [26] (MoS$_2$-MoTe$_2$), ref. [41] (WSe$_2$-Graphene). d, The gate modulation of RR at drain voltage of 0.52 V, 1V, 2V and 3V. e, The diagram of a half-wave rectifier circuit based on the heterostructure device. f, The half-wave rectifying results with input wave amplitude of 0.4 V and 2 V when the device gate voltage biased at 3 V and 0 V, respectively.

The linear scale plot of the $I_d$-$V_d$ curves are demonstrated in supplementary Figure 2. The turn on voltage of the diode can be efficiently tuned from around -2 V to -0.2 V by the applied gate voltage. Since the applied gate voltage will modulate the energy band of the WSe$_2$ layer, the slope of the turn on voltage as a function of the applied gate voltage can be regarded as gate-coupling efficiency.[20] The turn on voltage (defined as the V$_d$ where I$_d$= 1 nA) has a fast change with a gate-coupling efficiency of 0.71 when the gate voltage increases from -3 V to -1 V, and a slow change with a gate-coupling efficiency of 0.06 when the gate voltage increases from -0.5 V to 3 V. The main reason is that, when the gate voltage sweeps from -0.5 V to 3 V, the quasi Fermi level aligns near the conduction band edge of WSe$_2$, where a large density of states (DOS) exists.

The rectification ratio (RR) of the device versus drain bias under different gate voltages are shown in Figure 3b. The RR can be effectively modulated at a large scale of 7 orders by the applied gate voltage. Another WSe$_2$/SnSe$_2$ heterostructure device with WSe$_2$ and SnSe$_2$ thickness of 7.4 nm and 25.7 nm shown in supplementary Figure 3 presents an even higher RR modulation of 11 orders (from $10^{-5}$ to $10^6$) by gate voltage. The maximum realized RR and current density are more than $10^8$ and $10^4$ A/cm$^2$ at RT. To the best of our knowledge, these results are the highest reported values for semiconductor vdWH devices (See supplementary table 1 for more details). For comparison, the RRs of the devices from different vdWHs are ploted in Figure 3c. Due to the large band offset at the junction interface, the RR realized by WSe$_2$/SnSe$_2$ heterostructure here is much higher than other results.[16, 20, 26, 27, 41] Furthermore, the realized current density and RR in this work are comparable with the diodes based on conventional semiconductors (Table 1).

Table 1. Comparison of the room temperature figures of merit of the WSe2/SnSe2 diodes with the diodes based on conventional bulk semiconductors.

| Diode | Current density (A/cm$^2$) | Rectification ratio | Reference |
|---|---|---|---|
| WSe$_2$/SnSe$_2$ hetero diode | ~ **10$^4$** | ~ **10$^8$** | This work |
| GaAs-Si hetero diode | ~ 10$^4$ | ~ 10$^5$ | Scientific Reports 6, 25328 (2016) |
| CuO$_x$/InZnO$_x$ thin film diode | ~ 10$^4$ | ~ 10$^6$ | Advanced Materials 20, 3066-3069 (2008) |
| GaN p-n diode | ~ 10$^3$ | ~ 10$^7$ | Electronics Letters 52, 1170-1171 (2016) |
| Si thin film P-i-N diode | 10$^3$ | ~ 10$^7$ | Appl. Phys. Lett. 85, 2122-2124 (2004) |
| Si/SiGe resonant tunnel diode | ~ 10$^5$ | NA | Electron Device Letters 27, 364-367 (2006) |
| SiC P-i-N diode | ~ 10$^4$ | NA | IEEE transactions on magnetics 41(1), 330-333 (2005) |

The gate voltage dependent RR at drain voltages of 0.52 V, 1 V, 2 V and 3 V are demonstrated in Figure 3d. The RR shows a higher value at posive gate voltages. Even at a small drain bias of 0.52 V, the highest achieved RR is ~10$^5$ at V$_{bg}$= 1 V. As mentioned before, the turn on voltage of the diode can be as low as -0.2 V, which means a lower signal loss compared with conventional silicon based pn diode (turn on voltage is ~ 0.7 V) when used in a rectifier circuit. In a conventional small signal rectifier circuit, an amplifier is usually needed to amplify the signal first. Due to the high rectification ratio and low turn on voltage, the diode presented here can be used in small signal rectifier to simplify the circuit.[42] The device based half wave rectifier circuit and its performance are demonstrated in Figure 3e and f, respectively. Signals with amplitude of 0.4 V and 2 V are successfully rectified. The limited frequency of the rectifier is due to the large parasitic capacitances and large total resistance of the device (See supplementary Figure 5 for more details), which can be optimized by introducing a local gate device structure and increasing the area of the material. Since the nn backward diode presented here is conducted by majority carrier drifts, which do not suffer from the large minority carrier storage capacitance when forward biased in conventional diode, the device can also find great potential applications in high frequency switches, RF energy harvesting, and microwave detectors.[14, 43]

**Heterostructure device temperature dependent characterizations**

In order to better understand the carrier transport mechanism in the heterostructures, temperature dependent electrical characterizations of the $WSe_2/SnSe_2$ heterostructure device are carried out, using a device different from Figure 3. Since the pristine $WSe_2$ on $Al_2O_3$ substrate has a n-type dominated behavior, while the tunneling process happens during the $WSe_2$ is electrically p-type doped and the device is reverse biased, to enhance the BTBT process, $MoO_3$ coating is introduced to enhance the p-type behavior.[44] Shown in Supplementary Fig. 6 is the electrical behavior comparison of the device before and after $MoO_3$ coating. Since the work function (6.82 eV) of $MoO_3$ is more than 1 eV lower than the VBM of $WSe_2$, which facilitates the effective charge transfer between $MoO_3$ and $WSe_2$, the p-branch current of $WSe_2$ FET was increased by more than two orders of magnitude after $MoO_3$ coating. The doping concentration extracted from the transfer curves of $WSe_2$ FET is around $4.3 \times 10^{12}$ cm$^{-2}$ (supplementary Fig. 6b). When gate voltage biased at -5 V, the forward current of the $WSe_2/SnSe_2$ heterostructure device is increased by more than two orders of mignitude owing to the decreased hole injection barrier width between drain electrode and $WSe_2$ by $MoO_3$ doping (supplementary Fig. 6c and d).

Due to the unique properties, two dimensional material based heterostructures are regarded as a promising platform to realize TFETs with sub 60 mV/dec SS over a large current range.[45] Several reports have presented vdWHs based TFETs with sub 60 mV/dec SS in a limited current range.[18, 46] Figure 4a presents the room temperature $I_d$-$V_{bg}$ transfer curves of the $WSe_2/SnSe_2$ heterostructure device at drain bias of -0.5 V before and after $MoO_3$ coating. As expected, the on current of the p-TFET region increases around one order after $MoO_3$ coating. In this work, the minimum realized SS of the $WSe_2/SnSe_2$ heterostructure p-TFET is ~ 79 mV/dec (shown in supplementary Fig. 7), which is still above the thermal limit. Further methods to improve the gate efficiency and reduce the interface traps introduced during material transfer and device fabrication might help to reduce the SS. Instead of minimum SS, for practical application, a low average SS ($SS_{ave}$) over several decades of current change is more important to realize a low supply voltage. Thanks to the high gate coupling efficiency and low trap density (See supplementary Fig.8), $SS_{ave}$ of 126 mV/dec over five decades of current change at drain bias of -0.5 V is achieved in this work (See supplementary Fig. 9), which is a recording low $SS_{ave}$ value in vdWH based TFETs. This value is also comparable with the best reported p-TFET based on traditional semiconductor materials (Supplementary table 2). Considering the research of vdWHs based TFETs is still in very early stage, the promising application of vdWHs based TFETs in low power area can be anticipated in the near future.

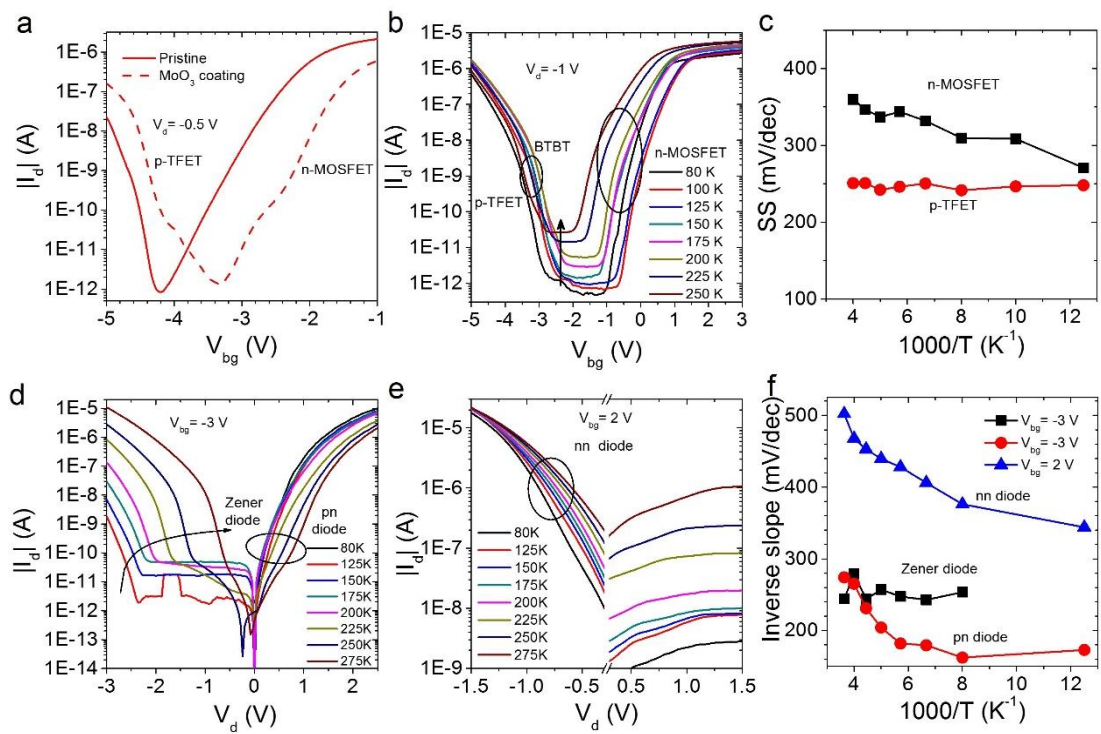

Figure 4. The doping of the heterostructures and temperature dependent electrical characterizations. a, Transfer curves of $WSe_2/SnSe_2$ heterostructure devices before and after 80 nm $MoO_3$ coating. Solid curve: before $MoO_3$ coating, dashed curve: after $MoO_3$ coating. b, The temperature dependent $I_d$-$V_{bg}$ transfer curves at $V_d$= -1 V. c, The SS of the p-TFET and n-MOSFET operated in the $WSe_2/SnSe_2$ heterostructure device as a function of 1000/T. d, Zener diode and pn diode $I_d$-$V_d$ curves at different temperatures when $V_{bg}$= -3 V. e, nn diode $I_d$-$V_d$ curves at different temperatures when $V_{bg}$= 2 V. f, The inverse slope of the pn diode, Zener diode and nn diode, operated in the $WSe_2/SnSe_2$ heterostructure, as a function of 1000/T.

The temperature dependent transfer curves of the $WSe_2/SnSe_2$ heterostructure devices at drain bias of -1 V are illustrated in Figure 4b. At large negative gate biases, the p-branch current with a relatively unchanged slope under different temperatures is observed, indicating the current is dominated by BTBT tunneling. Similar temperature independent current slope behavior was also observed in Si based TFETs.[47] The increment of the off state current (arrow indicated) with the temperature increasing is due to the trap assisted tunneling and the decrease of the tunneling band gap.[47-49] Meanwhile, the slope of the transfer curve at n-branch has a strong temperature dependence, which indicates the transport is dominated by thermal injection. Figure 4c shows the SS of the p-TFET and n-MOSFET, where the SS of p-TFET is almost temperature independent, while the SS of n-MOSFET increases as temperature increases. The relatively unchanged SS in the p-region under different temperatures confirms the tunneling operation mechanism of the $WSe_2/SnSe_2$ heterostructure device under this bias condition. Shown in supplementary Figure 10 is the transfer curve comparison between the $WSe_2$ FET and the $WSe_2/SnSe_2$ heterostructure device. When the gate voltage sweeps from positive to large negative voltage, the WSe$_2$ FET turns on earlier than the heterostructure device, indicating the WSe$_2$/SnSe$_2$ junction dominated the charge transport in the p-region, which is consistent with the temperature dependent measurement.

The output behavior of the device at different temperatures when $V_{bg}$=-3 V is investigated in Figure 4d. At negative drain bias region, as the temperature increased, the tunnel current onset voltage positively shifted, which is an indication of Zener tunneling[20]. The slope of the reverse current near the onset voltages remains relatively unchanged at different temperatures. While, the slope of the forward current near the onset voltage region decreases with elevated temperature. Shown in Figure 4e is the temperature dependent $I_d$-$V_d$ curve at $V_{bg}$=2 V. The carrier injection barrier at the Pt/WSe$_2$ interface decreases with increased temperature and the slope of the reverse current (the ellipse indicated region) of the nn backward diode shows a clear decrease with increasing temperature, indicating a thermal injection dominated process. At large negative bias ($V_d$ near -1.5 V), the current shows a relatively independent behavior with temperature, which implies the current is dominated by electron tunneling through the barrier. The forward current of the nn diode increases with elevated temperature, owing to the decreases of the barrier height at the WSe$_2$/SnSe$_2$ heterojunction interface and the temperature assisted electron tunneling through the barrier. The inverse slope of the Zener diode tunnel current, the pn diode forward current and the backward nn diode reverse current are shown in Figure 4f. The inverse slope of the Zener diode remains relatively unchanged due to its tunneling mechanism. While the inverse slope of the pn diode forward current and the nn backward diode reverse current increases with increasing temperature owing to their thermal injection process. These results are consistent with the temperature dependent transfer curve measurement and the band alignment assumption.

**Conclusions**

In summary, owing to the proper band alignment and material properties, we have demonstrated high performance van der Waals heterojunction devices composed of WSe$_2$ and SnSe$_2$. The electrical properties of the devices can be effectively tuned by the applied electric field, material thickness and doping. With these approaches, FET with current on/off ratio of more than $10^8$, diode with rectification ratio of more than $10^8$ and current density of more than $10^4$ A/cm$^2$, tunneling field effect transistor with SS$_{ave}$ of 126 mV/dec over 5 decades have been realized. Most of the values were much better than the results reported on other van der Waals heterostructures and comparable with traditional semiconductor devices. Temperature dependent electrical properties of the devices were also investigated, which confirmed the analysis of the transport mechanism. We believe, these results can contribute to pave the way for the commercial electronic device applications of van der Waals heterostructures.

**Methods**
**Device fabrication.** Heterostructures are prepared by mechanical exfoliation. The WSe$_2$ were exfoliated on 286 nm SiO$_2$/Si substrate, while SnSe$_2$ were exfoliated on 20 nm

$Al_2O_3$/Si substrate. First, $SnSe_2$ were picked up by using PPC or PMMA. Then placing the $SnSe_2$/PPC layer onto $WSe_2$ layer. Finally transfer the $WSe_2$/$SnSe_2$/PPC layer onto target substrate. The material exfoliate and transfer process were done in an argon filled glove box. Contact electrodes were fabricated by standard e-beam lithography and e-beam evaporation process.

**Electrical characterization.** The electrical properties were characterized in a vacuum probe station (Lakeshore, TTPX) at a pressure below $5 \times 10^{-6}$ mbar equipped with Keysight B1500A semiconductor analyser.

**Acknowledgements**

This work was supported by HiSilicon Research Department. W. L. acknowledges Cheng Zhou for the help on low temperature electrical measurements.

**Author contributions**

The manuscript was written through contributions of all authors. All authors have given approval to the final version of the manuscript.